\documentclass[manuscript]{aastex}
\shorttitle{Spatial and temporal distributions of THMF} 
\shortauthors{Ishikawa \& Tsuneta}

\begin{document}
\title{Spatial and temporal distributions of transient horizontal magnetic fields with deep exposure}
\author{Ryohko Ishikawa \altaffilmark{1,2} \and Saku Tsuneta\altaffilmark{2}}
\email{ryoko.ishikawa@nao.ac.jp}

\altaffiltext{1}{Department of Astronomy, University of Tokyo, Hongo,
Bunkyo-ku, Tokyo 113-0033, Japan} \altaffiltext{2}{National Astronomical
Observatory of Japan, 2-21-1 Osawa, Mitaka, Tokyo 181-8588, Japan}

\begin{abstract}
We obtained a long exposure vector magnetogram of the quiet Sun photosphere at the disk center with wide FOV of $51\arcsec \times 82\arcsec$. The observation was performed at \ion{Fe}{1}~525.0~nm with the shutter-less mode of the Narrow Band Filter Imager of the Solar Optical Telescope (SOT) on board \emph{Hinode} satellite. We summed the linear polarization ($LP$) maps taken with time cadence of 60~seconds for 2~hours to obtain a map with as long an exposure as possible. The polarization sensitivity would be more than 4.6 (21.2 in exposure time) times the standard observation with the SOT spectro-polarimeter. 
The $LP$ map shows a cellular structure with a typical scale of $5\arcsec - 10\arcsec$.
We find that the enhanced $LP$ signals essentially consist of the isolated sporadic transient horizontal magnetic fields (THMFs) with life time of 1-10 min, and are not contributed by long-duration weak horizontal magnetic fields. The cellular structure coincides in position with the negative divergence of the horizontal flow field, i.e., mesogranular boundaries with downflows. Azimuth distribution appears to be random for the scale size of the mesogranules. Some pixels have two separate appearances of THMFs, and the measured time intervals are consistent with the random appearance. THMFs tend to appear at the mesogranular boundaries, but appear randomly in time. We discuss the origin of THMFs based on these observations.
\end{abstract}

\section{Introduction}
The Spectropolarimeter (SP) of the Solar Optical Telescope \citep[SOT,][]{Tsuneta2008SoPh} on board \emph{Hinode} \citep{Kosugi2007} reveals ubiquitous granular-sized horizontal magnetic fields in the internetwork regions \citep{Orozco2007,Lites2007,Centeno2007}. Such horizontal fields are highly transient with lifetimes ranging from 1 to 10 minutes, and are called ''transient horizontal magnetic fields'' \citep[THMFs,][]{Ishikawa2008}. \citet{Ishikawa2010} showed that a THMF has the form of an isolated flux tube in an almost field-free photosphere using SIRGAUS inversion code \citep{Bellot2003}. 
These horizontal magnetic fields are observed in the quiet Sun, a weak plage region \citep{Ishikawa2008}, and the polar regions \citep{Tsuneta2008ApJ,Itoh2009}. The occurrence rate and magnetic field strength distribution (histogram) are exactly the same in these different regions in spite of a considerable difference in the amount of vertical magnetic flux \citep{Ishikawa2008,Itoh2009}. The mean field strength of THMFs is about 400~G \citep{Ishikawa2008}, which is comparable to the equipartition field strength corresponding to the granular motion. These results suggest that a local dynamo process due to the granular motion generates THMFs all over the Sun. 

\citet{Trujillo2004} observed that ubiquitous tangled magnetic fields with a mean field strength of $\sim100$~G exist on the spatial scale of the granulation (order of arcsec) by means of the Hanle effect. They pointed out that  magnetic fields detected with the Zeeman effect might be only the tip of the iceberg of the Sun's magnetism so that most of the magnetism is hidden from traditional observing methods. A fundamental question is whether most of the volume of the granular region is occupied by tangled magnetic fields or has sporadic distinct form of flux tubes.

In this Letter, we study temporal and spatial distributions of THMFs with deep exposure and with large field of view (FOV).
Such long-exposure polarization maps with very wide FOV allow us to obtain global distributions of much weaker horizontal magnetic fields in spatial and time domains. The spatial and temporal distributions would directly address the issues of the hidden magnetism \citep{Trujillo2004} and the origin of THMFs.

\section{Observation}
A quiet solar region at disk center was observed with a cadence of 60~s using the Narrowband Filter Imager (NFI) of SOT on 2009 April 25. The FOV is $51\arcsec \times 82\arcsec$, and the pixel size is $0\farcs16$ ($2\times$2 summed mode). 
We continuously take the Stokes $I$, $Q$, $U$, and $V$ images (in total 123 sets)  in the blue wing of the \ion{Fe}{1}~525.0~nm line with the exposure time of 4.8~sec (same as standard SP).  The spectral bandwidth of NFI is $\sim100$~m$\AA$. The dark current and the flat field corrections were performed. Residual crosstalk between the Stokes parameters was removed \citep{Ichimoto2008SoPh}. The observation was performed in the shutterless mode, whose modulation scheme is exactly the same as that of SP. The NFI noise level in the Stokes $Q$ and $U$ signals is measured by the standard deviation of each Stokes signals in pixels without any apparent polarization signal, and is $1.3\times10^{-3}I$. Since these pixels used for this measurement may contain weak polarization signals, the real noise level may be smaller than the number obtained here. 

We sum magnetograms over the whole time sequence. The exposures have constant cadences in 2 hours of observing time, and the total exposure time in the summed magnetograms is 9.8~min. This exposure time should be compared with 4.8~sec of the standard SP exposure time, and the nominal exposure time is a factor of 123 better in the summed NFI magnetogram. The noise level of the standard SP data is $1.2\times10^{-3}I$ for a single wavelength pixel \citep{Ishikawa2010}. For comparison, assuming that wavelength-integration for making polarization map is performed in the range of $\sim100$~m$\AA$ in the blue wing of \ion{Fe}{1}~630.25~nm line for SP, the noise level in the wavelength-integrated Stokes $Q$ and $U$ is $5.4\times10^{-4}I$. Thus, the signal to noise ratio of the summed NFI  magnetogram is a factor of 4.6  ($=\sqrt{123}\cdot 5.4\times10^{-4}/1.3\times10^{-3}$) better than that of the standard SP observation. Furthermore, the effective Lande g-factor is 3.0 for  the \ion{Fe}{1}~525.0~nm line, and 2.5 for the  \ion{Fe}{1}~630.25~nm line. The polarization sensitivity of the summed NFI magnetogram would be more than 4.6 (21.2 in exposure time) times the standard SP.

\begin{figure*}
\epsscale{0.6}
\plotone{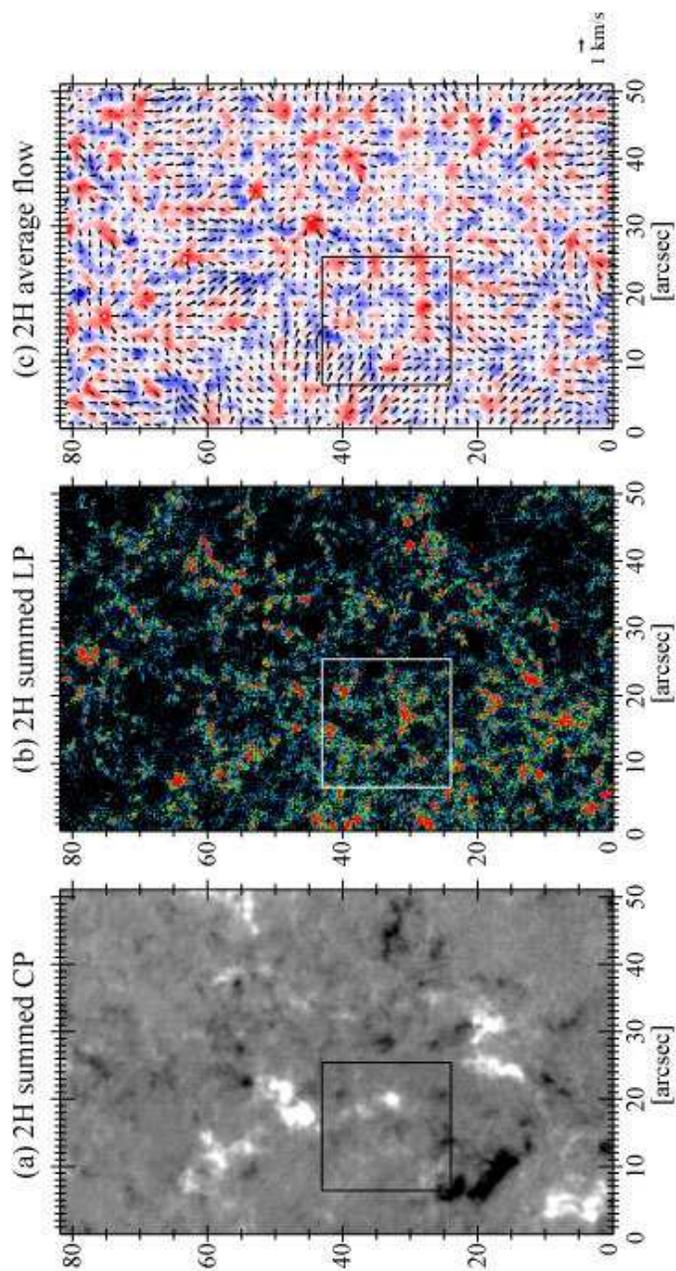}
\caption{(a) Two-hour summed $CP$ map. (b) Two-hour summed $LP$ map. Blue, green and red regions indicate statistical significance levels of $1\sigma$, $2\sigma$, and $3\sigma$ of the summed $LP$ map, respectively (see text). (c) Horizontal flow map plotted over the divergence map (the negative is blue and the positive red).}
\label{fig1}
\end{figure*}

\begin{figure*}
\epsscale{0.9}
\plotone{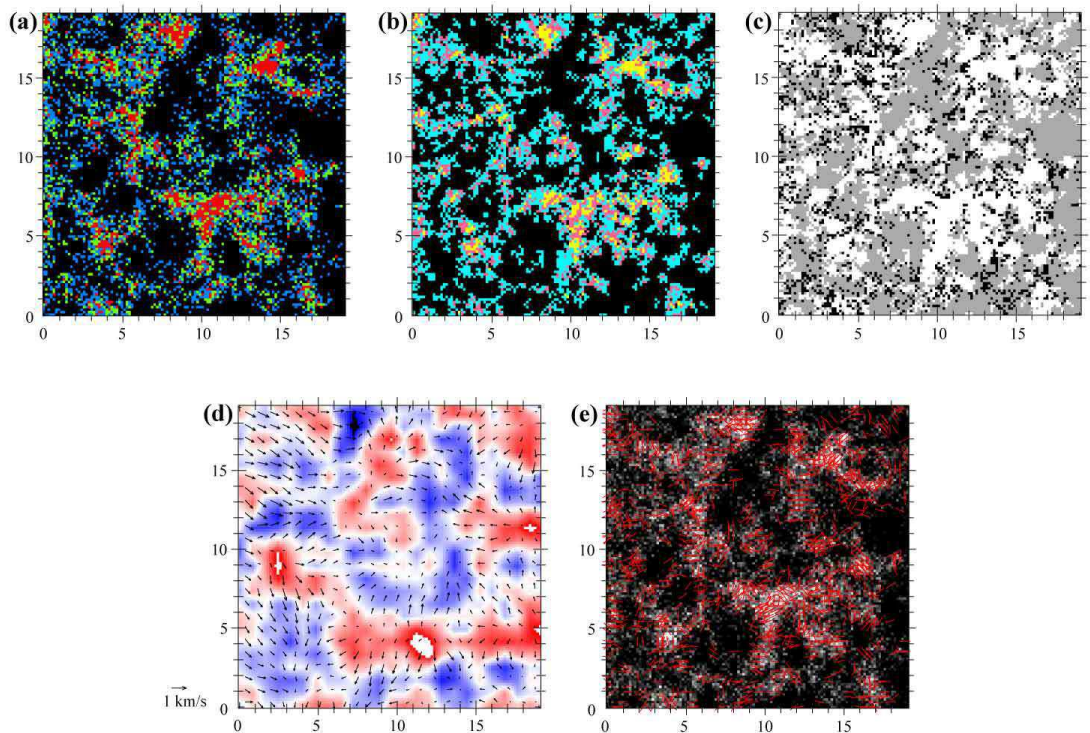}
\caption{All panels represent a small area of $19\farcs2\times19\farcs2$ (small box in Figure \ref{fig1}). (a) Summed $LP$ map. (b) Number of isolated THMF occurrence in time domain. Region where only one THMF appears during 2~hours are shown with light blue. Pink and yellow regions show regions where THMFs appear twice, and  more than twice, respectively. (c) The region with summed $LP>1\sigma$ \emph{and} with one or more isolated THMFs is shown with white, and the region with summed $LP>1\sigma$ \emph{and} without any apparent THMF is shown with black. (d) Horizontal flow map plotted over the divergence map (the negative is blue and the positive red). (e) Azimuth directions plotted over the summed $LP$ map. Original data have too many lines to show the azimuth directions, and  they are plotted every two pixels. }
\label{fig2}
\end{figure*}

\begin{figure*}
\epsscale{0.9}
\plotone{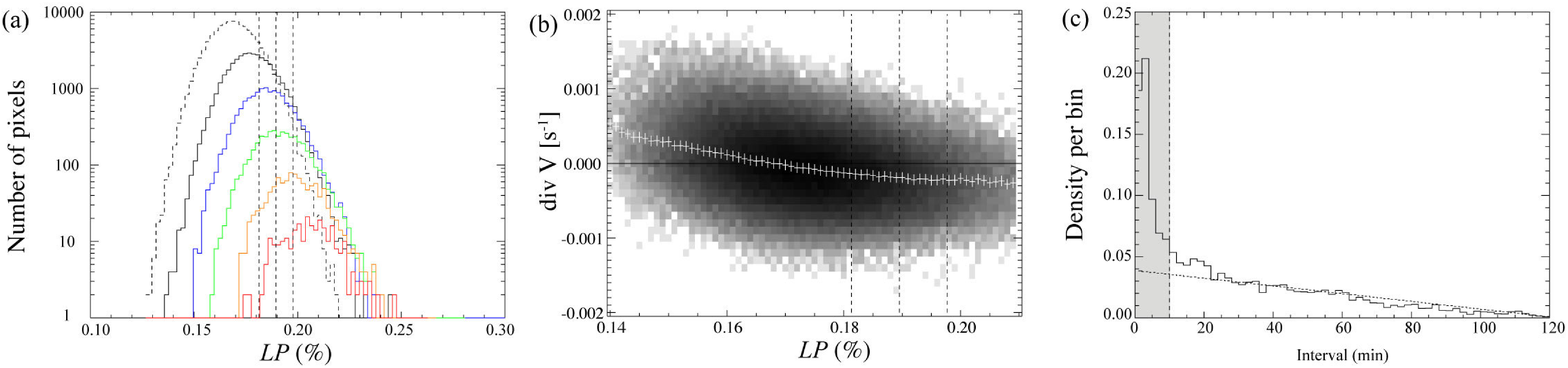}
\caption{(a) Histograms of the summed (long-exposure) $LP$ with different numbers of THMF occurrences. Solid black, blue, green, orange, and red lines represent the pixels where THMFs appear once, twice, three times, four times, five times, respectively. Dashed line corresponds to the pixels without any THMF during 2~hours of observation. (b) A scatter plot of the 2-hour summed $LP$ versus the divergence of the horizontal flow field. White crosses show the center of gravity of the distribution at each $LP$ bin. In (a) and (b), three dashed lines represent $1\sigma$, $2\sigma$, and $3\sigma$ noise level for the long-exposure $LP$ map. (c) Time interval distribution for pixels with THMFs appearing twice.}
\label{fig3}
\end{figure*}

\section{Deep linear polarization map}
\subsection{Spatial distribution of linear polarization}
Figure~\ref{fig1} (a) shows the map of the circular polarization ($CP$) summed over the whole observing duration. The FOV has a size that approximately covers one supargranular cell, which is delineated by the scattered vertical network fields.
We sum 123 maps of linear polarization, $LP=\sqrt{Q^{2}+U^{2}}/I$ (Figure \ref{fig1} (b)). The deep $LP$ map should potentially show spatial distribution of much  weaker horizontal magnetic fields, which have not been detected in the single $LP$ map. The summed $LP$ map clearly shows cellular structures on a scale of $5\arcsec - 10\arcsec$ with voids.  

To evaluate the statistical significance of these apparent structures, we here estimate the noise level of the summed $LP$ map. We derive the standard deviation, $\sigma_{0}$ ($\sigma_{0}=1.0\times10^{-3}$) of the photon noise of a single $LP$ map from the  pixels without any enhanced linear polarization signals. The standard deviation, $\sigma$ of the summed $LP$ map is then simply given by $\sigma=\sigma_{0}/\sqrt{n}$, where $n$ is the number of the $LP$ images (n=123). In Figure \ref{fig1}~(b), the regions with summed $LP$ above $1\sigma$, $2\sigma$, and $3\sigma$ levels are shown. 30\% of the FOV possess $LP>1\sigma$. Figure \ref{fig1}~(b) clearly shows that the cellular structure of the horizontal fields with voids is statistically significant.

\subsection{Occurrence map of THMFs}
\label{occurrence}
Does the cellular distribution in Figure\ref{fig1}~(b) consist of  isolated sporadic THMFs or weaker magnetic fields, which persist longer? To answer this question, we pick up THMFs from the individual $LP$ maps. The appearance of a THMF is defined by a timing when the area with $LP>3\sigma_{0}$ becomes larger than 4~pixels. We track the thus-identified THMF as long as $LP>3\sigma_{0}$. Except for the areal threshold of 4~pixels used to identify the start of an event, there is no area threshold during the subsequent tracking. 37\% of the FOV in the summed map are pixels where THMFs occurred over the course of the observations. 

We also investigate how many THMFs appear in the same pixel during the 2~hours. In 67~\% of the pixels that possess at least one THMF, THMFs appear once, and in 24~\% THMFs  appear twice. In this case, two THMFs separated in time appear during the observing time of 2 hours. Figure~\ref{fig2} (b) represents a number of the occurrence in the small area indicated with the box in Figure~\ref{fig1}. The black area corresponds to the region where no THMFs appear during 2~hours.  In the occurrence map (Figure~\ref{fig2}~(b)), the cellular structure on a scale of $\sim10\arcsec$ is identified and the structure is similar to that of the summed $LP$ map in Figure~\ref{fig2}~(a).

Figure~\ref{fig3}~(a) shows histograms of the 2-hour summed $LP$ map. Each line represents the $LP$ distribution for pixels with different numbers of THMF appearances. 
Here we focus on the summed $LP$ histogram for pixels without any THMF appearance shown with dashed line. In spite of no appearance of THMFs, 16\% of such pixels possesses summed $LP$ greater than 1$\sigma$ (4\% greater than $2\sigma$). Do they represent the hidden magnetic fields \citep{Trujillo2004}? In Figure~\ref{fig2}~(c), the pixels with the summed $LP>1\sigma$ \emph{and} without any THMF appearance for 2~hours are shown with black, while the pixels with the summed $LP>1\sigma$ \emph{and} with one or more THMF appearance are shown with white. We notice that the black dot-like regions clearly delineate the boundaries of the white regions, where THMFs appear at least once. This indicates that these black pixels are also occupied by THMFs, but their $LP$ does not reach the $3\sigma_{0}$ threshold. In the summed $LP$ map, we do not find any signature of weak extended horizontal magnetic fields. Instead, the structure seen in the deep exposure map consists of individual isolated THMFs.

The peak of the histogram in Figure~\ref{fig3}~(a) is shifted toward higher $LP$ with an increasing number of THMF appearances. This indicates that the regions with multiple appearances of THMFs have higher $LP$. Examples of such multiple appearances are found at positions [$10\arcsec$, $6\arcsec$] and [$14\arcsec$, $15\arcsec$] in Figure~\ref{fig2} (b). On the other hand, the number of occurrences is not necessarily high in the pixels even with high summed $LP$. This is consistent with the fact that the $LP$ histograms overlap with $LP>0.22\%$ regardless of the different number of THMF occurrences (Figure~\ref{fig3} (a)). For instance, at  [$7\arcsec$, $10\arcsec$] in Figure~\ref{fig2} (b), even though  2-hour summed $LP$ is high, the number of occurrences is 1. In such regions, THMFs with long lifetimes or with higher linear polarization signals would appear. We confirm that regions with statistically significant summed $LP$ essentially have an isolated THMFs at least once for two hours of observation time.

\subsection{Relation with photospheric horizontal flow}
\label{meso}
In the previous two sections, we show that the regions where THMFs appear have structures on a scale of $5\arcsec-10\arcsec$, which is larger than a typical granule size. In this section we compare the summed $LP$ map with the  photospheric flow pattern with a spatial scale larger than that of granules.
We apply a subsonic filter to the time series of Stokes $I$ images with a cut off velocity at 10~km~s$^{-1}$ \citep{Title1989}. From these filtered images, we obtain the horizontal flow velocity using the local correlation technique \citep[e.g.,][]{November1988,Shine2000}.  This technique compares the small subfields in each image with those in the subsequent image, and obtains the displacement that provides the best correlation. The small subfields are defined by the apodization window with a Gaussian function. The choice of the FWHM of the apodization window should depend on the spatial scale of what we want to track. For the present work we use an FWHM of  $1\farcs6$ \citep{Okamoto2009}, and derived velocity maps are averaged over the whole time sequence to measure the horizontal flow with scale size larger than the granular flow in temporal and spatial domain. 

Figure~\ref{fig1}~(c) represents the horizontal flow field plotted over the divergence map ($\nabla_{\parallel} \cdot v_{\parallel}=\frac{\partial v_{x}}{\partial x}+\frac{\partial v_{y}}{\partial y}$). The positive divergence is distributed with a scale of $5\arcsec - 10\arcsec$, and corresponds to mesogranules \citep[e.g.,][]{Shine2000}. Comparing Figure~\ref{fig1}~(b) with (c), it appears that the $LP$ avoids the region with the positive divergence (see movie). 
Figure~\ref{fig3}~(b) represents the relationship between the $LP$ signals summed over  2~hours and the divergence of the horizontal flow field. We find a weak dependence of $LP$ on the divergence: Pixels with weak $LP$ signals appear to have positive divergence, while pixels with strong $LP$ signals appear to have negative divergence. The $LP$ signals are essentially concentrated on the negative divergence area,  which corresponds to the boundary of mesogranules. On the other hand, we do not find any relationship of the summed $LP$ with the rotation of the horizontal flow field.

\subsection{Azimuth distribution}
In Figure~\ref{fig2} (e) we plot the azimuth directions of THMFs when the area of each THMF become the largest through their time evolution. When multiple THMFs appear at different times at the same position, the lines showing the azimuth direction of each THMF overlap. We find no coherent magnetic field lines extending to a mesogranular scale. Furthermore, even though THMFs appear at several times, the azimuth directions seem to be random. These results would suggest that there are no coherent magnetic fields on a mesogoranular scale lying below the solar surface, or even if such magnetic fields exist, the field strength is too weak to maintain its memory to the surface.

\subsection{Temporal distribution}
In this section we examine whether there is any  typical timescale for the THMF appearance. As we described in Section \ref{occurrence}, 37\% of the pixels throughout the FOV possess THMFs, and THMFs appear twice  in 24\% of the pixels, which have THMFs at least once. Figure~\ref{fig3}~(c), solid line represents time intervals between the first and the second THMF appearances at each pixel in which THMF appears twice for 2~hours. We find a strong peak at 2-4 minutes.  THMFs have lifetimes ranging from 1~min to 10~min \citep{Ishikawa2008}, and this peak would originate from the false identification of THMFs with $LP$ close to the detection threshold. We normalize the histogram by the total number of pixels with time intervals of two THMFs longer than 10~minutes. Over 10~minutes, the distributions do not show any peak. For comparison, as summarized by \citet{Leitzinger2005}, the lifetime of the mesogranules ranges from 30~min to 6~hours  depending on analyses, while the turnover time of granules 
is believed to be of the order of 1000~sec. Though our observing time is limited to 2 hours, we do not see any signature associated with these time scales of the flow field.

Dashed lines show the distribution of time intervals when THMFs appear randomly twice for 2~hours. To put 2 balls (THMFs) into 123 linearly numbered boxes (time bins) ${}_{123}\mathrm{C}_{2}$=7503 combinations are possible in total, and  we measure the intervals of two balls for all cases. This corresponds to the case of random distribution (dashed line in Figure~\ref{fig3}~(c)). There is a remarkable coincidence between the measured interval of THMFs and the interval of the  random distribution. This shows that the multiple appearances of THMF are independent, i.e. random, and are not causally related. Appearances of THMFs are organized by mesogranular velocity field in spatial domain, but are random in time domain.

\section{Discussions}
We have analyzed the vector magnetograms taken at \ion{Fe}{1}~525.0~nm with long exposure time to investigate spatial and temporal distributions of THMFs. We find that the deep-exposure $LP$ map has a clear cellular structure with spatial scale of $5\arcsec - 10\arcsec$. Although we anticipate with its better signal to noise ratio that such an $LP$ map allows us to detect weak polarization signals possibly extended in temporal and spatial domains, we do not find any such structure. Instead, we demonstrate in this letter that such a summed $LP$ map essentially represents the occurrence distribution of THMFs, which we see in much shorter exposure $LP$ maps. We infer that, in the quiet Sun, isolated horizontal flux tubes exist in the essentially field-free regions. Such isolated flux tubes are sporadically (intermittently) observed as THMFs in the photosphere.

We compare the deep-exposure $LP$ map with the photospheric horizontal flow field, and find a correlation between the flow field and the distribution of $LP$: The regions abundant in $LP$ coincide in position with negative divergence areas. Since $\nabla \cdot v=\nabla_{\parallel} \cdot v_{\parallel}+\nabla_{\perp} \cdot v_{\perp}$=0, this negative divergence area is the region with downflow, corresponding to the boundary of the mesogranules. Interestingly, THMFs do not appear in the intergranular lanes where downflow dominates \citep{Centeno2007,Ishikawa_Hinode2}. The location of THMFs is restricted to the downflow region of mesogranules, although the appearance of THMFs would be driven by the granular upward motion.

The azimuth directions of THMFs appear to be random on a mesogranular scale. When THMFs appear more than once, their azimuth directions and the time interval are random. The random appearance of THMFs in time may not be consistent with a scenario of observed horizontal magnetic fields simply being maintained by recycling (without any dynamo process) of magnetic fields in the granular convective motions: the emerged horizontal magnetic fields do not come back to the convection zone, and are not subject to recycling and subsequent emergence. Instead, they reach above the photosphere as suggested by \citet{MartinezGonzalez2009} and \citet{Ishikawa2010}. 

The mesogranular pattern was also found for vertical magnetic elements \citep[e.g.,][]{deWijn2005}. \citet{Lites2007} reported from a single SOT/SP snap shot that linear polarization signal as well as circular one are organized on a scale of  $5\arcsec - 10\arcsec$. Are both the vertical magnetic fields and THMFs transported to the mesogranular boundaries with the same mechanism? Such a concentrated circular polarization signal may be at least partially contributed by the footpoints of THMFs found in the mesogranular boundaries. \citet{Dominguez2003} simulated the motion of corks advected by the observed flow velocity field, and found that the observed vertical magnetic elements have a distribution similar to that of the corks (mesogranular concentration). One difference between the two cases (vertical vs. horizontal) is that the vertical magnetic fields tend to be pinned down and can stay in the granular and mesogranular downflow region, while the horizontal fields may be subject to the continuous turnover motion.  Indeed, no THMFs appear in the intergranular lanes, while we find elemental vertical flux tubes in the intergranular lanes.  

If THMFs are transported by the mesogranular flow and somehow stay in the mesogranular downflows, these THMFs should have had more uniform azimuthal distribution. In fact, there is no correlation in the azimuthal angles of THMFs. The mean field strength of THMFs is comparable to the equipartition field strength ($B_{eq}$) corresponding to the granular motion, and is much stronger than $B_{eq}$ corresponding to mesogranular flow velocity of 300 - 500~m~s$^{-1}$ \citep[see Table 1 in][]{Leitzinger2005}. This may make the transport of the THMFs with mesogranular flow difficult. These considerations indicate that THMFs are not advected like the vertical magnetic fields.

Then, why do we see such clear concentrations of THMFs around the boundaries of mesogranules? Any dynamo mechanism needs seed magnetic fields. Such seed fields with field strength less than the mesogranular $B_{eq}$ may be easily advected to the boundaries of mesogranulation, and a local dynamo mechanism amplifies the field strength from mesogranular $B_{eq}$ to granular $B_{eq}$. The local dynamo process can occur anywhere due to the ubiquitous granular motion. But, the seed fields are provided to the mesogranular boundaries by the mesogranular flow field.

As a last note, we have not discussed something important, namely the relationship between THMFs and supergranulation. The lack of the observations is due to the limitation of the FOV in this particular observation (Figure~\ref{fig1} (a)). In order to obtain the supergranular flow pattern by smoothing out the granulation and the mesogranulation, a longer observation and a larger FOV is necessary \citep{Shine2000}. This is left to future Hinode and/or groundbased observations. 

\acknowledgements 
Our observation belongs to the Hinode Operation Plan (HOP) 71, supported by K. Ichimoto, R. A. Shine, and T. J. Okamoto.
The authors thank V. Mart{\'i}nez Pillet for stimulating discussion.
\emph{Hinode} is a Japanese mission developed and launched by ISAS/JAXA, with NAOJ as a domestic partner and NASA and STFC (UK) as international partners. It is operated by these agencies in co-operation with ESA and NSC (Norway).


\begin{thebibliography}{}
\bibitem[Bellot Rubio(2003)]{Bellot2003} Bellot Rubio, L.~R.\ 
2003, Astronomical Society of the Pacific Conference Series, 307, 301 

\bibitem[Centeno et al.(2007)]{Centeno2007} Centeno, R., et al.\ 
2007, \apjl, 666, L137 

\bibitem[{de Wijn}(2005){de Wijn}, {Rutten}, {Haverkamp}, {S{\"u}tterlin}]{deWijn2005}
{de Wijn}, A.~G. et al. 2005, \aap, 441, 1183

\bibitem[{Dom{\'{\i}}nguez Cerde{\~n}a}(2003){Dom{\'{\i}}nguez Cerde{\~n}a}]{Dominguez2003}
{Dom{\'{\i}}nguez Cerde{\~n}a}, I. 2003, \aap, 412, L65

\bibitem[{{Ichimoto} {et~al.}(2008){Ichimoto}, {Lites}, {Elmore}, {Suematsu},
  {Tsuneta}, {Katsukawa}, {Shimizu}, {Shine}, {Tarbell}, {Title}, {Kiyohara},
  {Shinoda}, {Card}, {Lecinski}, {Streander}, {Nakagiri}, {Miyashita},
  {Noguchi}, {Hoffmann}, \& {Cruz}}]{Ichimoto2008SoPh}
{Ichimoto}, K., {Lites}, B., {Elmore}, D., {et~al.} 2008, \solphys, 249, 233

\bibitem[{{Ishikawa} \& {Tsuneta}(2009{\natexlab{a}})}]{Ishikawa2009}
{Ishikawa}, R. \& {Tsuneta}, S. 2009{\natexlab{a}}, \aap, 495, 607

\bibitem[{{Ishikawa} \& {Tsuneta}(2009{\natexlab{b}})}]{Ishikawa_Hinode2}
{Ishikawa}, R. \& {Tsuneta}, S. 2009{\natexlab{b}}, Hinode 2 ASP Conference Series, in press

\bibitem[{{Ishikawa}, {Tsuneta}, \& {Jur\v{c}{\'a}k}(2010)}]{Ishikawa2010}
{Ishikawa}, R., {Tsuneta}, S., \& {Jur\v{c}{\'a}k}, J. 2010, \apj, in press

\bibitem[{{Ishikawa} {et~al.}(2008){Ishikawa}, {Tsuneta}, {Ichimoto}, {Isobe},
  {Katsukawa}, {Lites}, {Nagata}, {Shimizu}, {Shine}, {Suematsu}, {Tarbell}, \&
  {Title}}]{Ishikawa2008}
{Ishikawa}, R., {Tsuneta}, S., {Ichimoto}, K., {et~al.} 2008, \aap, 481, L25

\bibitem[Ishikawa et al.(2010)]{Ishikawa2010} Ishikawa, R., Tsuneta, 
S., \& Jur{\v c}{\'a}k, J.\ 2010, \apj, 713, 1310 

\bibitem[{{Ito} {et~al.}(2010){Ito}, {Tsuneta}, {Shiota}, {Tokumaru}, \&
  {Fukui}}]{Itoh2009}
{Ito}, H., {Tsuneta}, S., {Shiota}, D., {Tokumaru}, M., \& {Fukui}, K. 2009,
  \apj, submitted

\bibitem[{{Kosugi} {et~al.}(2007){Kosugi}, {Matsuzaki}, {Sakao}, {Shimizu},
  {Sone}, {Tachikawa}, {Hashimoto}, {Minesugi}, {Ohnishi}, {Yamada}, {Tsuneta},
  {Hara}, {Ichimoto}, {Suematsu}, {Shimojo}, {Watanabe}, {Shimada}, {Davis},
  {Hill}, {Owens}, {Title}, {Culhane}, {Harra}, {Doschek}, \&
  {Golub}}]{Kosugi2007}
{Kosugi}, T., {Matsuzaki}, K., {Sakao}, T., {et~al.} 2007, \solphys, 118

\bibitem[{Leitzinger}{et~al.}(2005){Leitzinger}, {Brandt}, {Hanslmeier}, {P{\"o}tzi}, \&{Hirzberger}]{Leitzinger2005}
{Leitzinger}, M. et al. 2005, \aap, 444, 245

\bibitem[{{Lites} {et~al.}(2008){Lites}, {Kubo}, {Socas-Navarro}, {Berger},
  {Frank}, {Shine}, {Tarbell}, {Title}, {Ichimoto}, {Katsukawa}, {Tsuneta},
  {Suematsu}, {Shimizu}, \& {Nagata}}]{Lites2007}
{Lites}, B.~W., {Kubo}, M., {Socas-Navarro}, H., {et~al.} 2008, \apj, 672, 1237

\bibitem[{{Mart{\'{\i}}nez Gonz{\'a}lez} \& {Bellot
  Rubio}(2009)}]{MartinezGonzalez2009}
{Mart{\'{\i}}nez Gonz{\'a}lez}, M.~J. \& {Bellot Rubio}, L.~R. 2009, \apj, 700,
  1391

\bibitem[{{November} \& {Simon}(1988){November}, \& {Simon}}]{November1988}
{November}, L.~J. and {Simon}, G.~W. 1988, \apj, 333, 427

\bibitem[{{Okamoto} {et~al.}(2009){Okamoto}, {Tsuneta}, {Lites}, {Kubo}, {Yokoyama},  {Berger}, {Ichimoto}, {Katsukawa}, {Nagata}, {Shibata}, {Shimizu}, {Shine}, {Suematsu}, {Tarbell}, \&{Title}}]{Okamoto2009}
{Okamoto}, T.~J. {et~al.} 2009, \apj, 697, 913

\bibitem[{{Orozco Su{\'a}rez} {et~al.}(2007){Orozco Su{\'a}rez}, {Bellot
  Rubio}, {del Toro Iniesta}, {Tsuneta}, {Lites}, {Ichimoto}, {Katsukawa},
  {Nagata}, {Shimizu}, {Shine}, {Suematsu}, {Tarbell}, \& {Title}}]{Orozco2007}
{Orozco Su{\'a}rez}, D. {et~al.} 2007{\natexlab{a}}, \apjl, 670, L61


\bibitem[{{Shine} {et~al.}(2000){Shine}, {Simon}, \& {Hurlburt}}]{Shine2000}
 {Shine}, R.~A. and {Simon}, G.~W. and {Hurlburt}, N.~E. 2000, \solphys, 193, 313


\bibitem[Title et al.(1989)]{Title1989} Title, A.~M., Tarbell, 
T.~D., Topka, K.~P., Ferguson, S.~H., Shine, R.~A., 
\& SOUP Team 1989, \apj, 336, 475 


\bibitem[Trujillo Bueno et al.(2004)]{Trujillo2004} Trujillo Bueno, 
J., Shchukina, N., \& Asensio Ramos, A.\ 2004, \nat, 430, 326 
  
\bibitem[{{Tsuneta} {et~al.}(2008{\natexlab{a}}){Tsuneta}, {Ichimoto},
  {Katsukawa}, {Nagata}, {Otsubo}, {Shimizu}, {Suematsu}, {Nakagiri},
  {Noguchi}, {Tarbell}, {Title}, {Shine}, {Rosenberg}, {Hoffmann}, {Jurcevich},
  {Kushner}, {Levay}, {Lites}, {Elmore}, {Matsushita}, {Kawaguchi}, {Saito},
  {Mikami}, {Hill}, \& {Owens}}]{Tsuneta2008SoPh}
{Tsuneta}, S. {et~al.} 2008{\natexlab{a}},
  \solphys, 249, 167

\bibitem[{{Tsuneta} {et~al.}(2008{\natexlab{b}}){Tsuneta}, {Ichimoto},
  {Katsukawa}, {Lites}, {Matsuzaki}, {Nagata}, {Orozco Su{\'a}rez}, {Shimizu},
  {Shimojo}, {Shine}, {Suematsu}, {Suzuki}, {Tarbell}, \&
  {Title}}]{Tsuneta2008ApJ}
{Tsuneta}, S. {et~al.} 2008{\natexlab{b}},
  \apj, 688, 1374

\end{thebibliography}
\end{document}